\documentclass[iop]{emulateapj}

\usepackage[colorlinks=true,linkcolor=blue,anchorcolor=blue,citecolor=blue,urlcolor=blue]{hyperref}  
\hypersetup{pdfauthor={Ho Seong HWANG}, pdftitle={The SMA survey of Local DOGs}}

\newcommand{\iras}{\textit{IRAS} }

\newcommand{\wise}{\textit{WISE} }

\newcommand{\spitzer}{\textit{Spitzer} }

\slugcomment{Last updated: \today}
\shorttitle{The SMA survey of Local DOGs}
\shortauthors{Hwang et al.}

\usepackage{natbib}

\begin{document}

\title{DUST PROPERTIES OF LOCAL DUST-OBSCURED GALAXIES WITH THE SUBMILLIMETER ARRAY}

\author{Ho Seong Hwang, Sean M. Andrews, and Margaret J. Geller}
\affil{Smithsonian Astrophysical Observatory, 60 Garden Street, Cambridge, MA 02138, USA}
\altaffiltext{1}{E-mail: hhwang@cfa.harvard.edu, sandrews@cfa.harvard.edu, mgeller@cfa.harvard.edu}


\begin{abstract}

We report Submillimeter Array (SMA) observations of the 880 $\mu$m
  dust continuum emission for 
  four dust-obscured galaxies (DOGs) in the local universe.
Two DOGs are clearly detected with $S_\nu$(880 $\mu$m) $=10-13$ mJy and $S/N>5$,
  but the other two are not detected 
  with 3$\sigma$ upper limits of $S_\nu$(880 $\mu$m) $=5-9$ mJy.
Including an additional two local DOGs with submillimeter data from the literature,
  we determine the dust masses and temperatures
  for six local DOGs.
The infrared luminosities and dust masses for these DOGs
  are in the range $1.2-4.9\times10^{11} (L_\odot)$
  and $4-14\times10^{7} (M_\odot)$, respectively.
The dust temperatures derived from 
  a two-component modified blackbody function
  are $23-26$ K and $60-124$ K
  for the cold and warm dust components, respectively.
Comparison of local DOGs with other infrared luminous galaxies
  with submillimeter detections
  shows that the dust temperatures and masses
  do not differ significantly among these objects.
Thus, as argued previously, local DOGs are not a distinctive population
  among dusty galaxies, but 
  simply represent the high-end tail of the dust obscuration distribution.
\end{abstract}


\section{INTRODUCTION}

Recent studies suggest that
  the cosmic star formation density peaks around $z=2$,
  and then decreases by an order of magnitude towards $z=0$
  (e.g., \citealt{bmag13,beh13}).
Interestingly, over the last 11 billion years,
  this cosmic star formation density is dominated by
  infrared luminous galaxies rather than ultraviolet (UV) luminous galaxies
  \citep{tak05,red08, red12,bou10,hei13,bur13}.
Therefore, studying high-$z$ dusty galaxies
  is critical for understanding the change in 
  the star formation activity of galaxies
  with cosmic time \citep{elb11,lutz11,oli12}.
  
Among many methods for identifying high-$z$ dusty galaxies,
  a simple optical/mid-infrared color criterion
  with ($R-[24]$) $\geq 14$ (mag in Vega, or
  $S_\nu$(24 $\mu$m)/$S_\nu$($R$)$\geq982$)
  is very efficient in selecting 
  $z\sim2$ star-forming galaxies 
  with large dust obscuration:
  dust-obscured galaxies (DOGs, \citealt{dey08,fio08,pen12,hwa12shels}).

These DOGs seem responsible for 10--30\%
  of the total star formation rate density
  of the universe at $z=1.5-2.5$ \citep{cal13}.  
These objects are divided into two groups depending on the shape
  of their spectral energy distributions (SEDs) 
  at rest-frame near- and mid-infrared wavelengths:
  ``bump'' and ``power-law'' DOGs \citep{dey08}.
The SEDs of bump DOGs show a rest-frame 1.6 $\mu$m stellar bump,
  resulting from the minimum opacity of the H$^{-}$ ion
  in the atmospheres of cool stars \citep{john88}.
In contrast, the power-law DOGs have a rising continuum
  with weak polycyclic aromatic hydrocarbon (PAH) emission,
  probably resulting from the hot dust component 
  heated by active galactic nucleus (AGN) \citep{hou05,desai09}.

Numerical simulations suggest that
   the DOGs are a diverse population ranging from
   intense gas-rich galaxy mergers to
   secularly evolving star-forming disk galaxies \citep{nar10}.
However, because of their extreme distances,
  it is difficult to fully understand the nature of these extremely dusty galaxies.

To study the physical properties of DOGs in detail
  (e.g., morphology, SED, dust mass and temperature),
  we focus on the rare local analogs of DOGs discovered recently 
  (\citealt{hg13dog}, hereafter HG13).
Thanks to their proximity and the wealth of multiwavelength data,
  the local DOGs are a useful testbed for studying what makes a DOG a DOG
  and for improving the understanding of the nature of their high-$z$ siblings.  

Using the {\it Wide-field Infrared Survey Explorer} ({\it WISE}; \citealt{wri10}) and 
  {\it Galaxy Evolution Explorer} ({\it GALEX}; \citealt{mar05}) data,
  we identified 47 DOGs at $0.05<z<0.08$ with large flux density ratios
  between mid-infrared ({\it WISE} 12 $\mu$m) and 
  near-UV ({\it GALEX} 0.22 $\mu$m) bands\footnote{We first 
  used {\it AKARI} 9 $\mu$m and {\it GALEX} NUV data,
  roughly equivalent to
  the $R$-band (0.65 $\mu$m) and \spitzer 24 $\mu$m data 
  originally used for selecting $z\sim2$ DOGs \citep{dey08}.
  We then used {\it WISE} 12 $\mu$m data instead of 
  {\it AKARI} 9 $\mu$m to increase the sample size (see HG13 for details).}
  [i.e., $S_\nu$(12 $\mu$m)/$S_\nu$(0.22 $\mu$m)$\geq 892$]
  in the Sloan Digital Sky Survey (SDSS, \citealt{york00})
  data release 7 (DR7, \citealt{aba09}).
The observational data for local and high-$z$ DOGs
  suggest a common underlying physical origin of the two populations;
  both seem to represent the high-end tail of the dust obscuration distribution 
  resulting from various physical mechanisms
  rather than a unique phase of galaxy evolution (HG13).

\begin{deluxetable*}{ccccclrrc}
\tabletypesize{\footnotesize}
\tablewidth{0pc} 
\tablecaption{SMA Observing Journal
\label{tab-samp}}
\tablehead{
ID\tablenotemark{a} & SDSS ObjID (DR9) & R.A.$_{2000}$ & Decl.$_{2000}$ & z & UT Date & $S_{880 \mu m}$ & $\sigma$ & Other Name\\
        &                  &               &                &   &         & (mJy)           & (mJy)    & 
	}
\startdata
 LDOG-07 &    1237674462024106294 & 09:04:01.02 & $+$01:27:29.12 & 0.0534 & 2013 Feb 12 & 12.5  & 1.4 &  \\
 LDOG-26 &    1237667209992732748 & 12:21:34.35 & $+$28:49:00.12 & 0.0613 & 2013 Feb 12 & 10.2  & 2.0 &  \\
 LDOG-39 &    1237648705135051235 & 15:26:37.67 & $+$00:35:33.50 & 0.0507 & 2013 Apr 1  & $<$4.8 & ... & CGCG 021-096 \\ 
 LODG-41 &    1237662663216070833 & 15:51:53.04 & $+$27:14:33.65 & 0.0589 & 2013 Apr 1  & $<$8.9 & ... & \\
\hline
 LDOG-08 &    1237674460413690092 & 09:07:46.91 & $+$00:34:30.55 & 0.0534 & ...         & ...  & ... & \\
 LDOG-35 &    1237665430243704865 & 14:07:00.39 & $+$28:27:14.67 & 0.0770 & ...         & ...  & ... & MRK 668
\enddata
\tablenotetext{1}{IDs in Table 1 of HG13.}
\end{deluxetable*}

The current multiwavelength data for local DOGs
  mostly cover only $\lambda\leq100$ $\mu$m 
  from the {\it Infrared Astronomical Satellite} ({\it IRAS}; \citealt{neu84});
  there are only five DOGs with {\it AKARI} 140 $\mu$m data \citep{mur07}.
There are no useful data on the `Rayleigh-Jeans' side of the infrared SED peak;
  these data are essential for deriving dust temperatures and 
  dust masses for these galaxies
  \citep{hwa10tdust,dale12,sym13}.
Quantifying the dust properties is important because
  the combination of dust and stellar properties gives better constraints 
  on the nature of these heavily obscured galaxies.
We can also directly compare these quantities 
  with model predictions \citep{nar10}.
The comparison of these local DOGs with other dusty galaxies 
  can establish a possible evolutionary link among them.

We thus conducted Submillimeter Array (SMA; \citealt{ho04})
 observations of the 880 $\mu$m continuum emission 
 for four bright local DOGs to derive the physical parameters of their dust content. 
We report the results from this pilot survey.
Section \ref{data} describes the sample and 
  the details of the SMA observations and data reduction.
We derive the physical parameters of the dust content in local DOGs,
  and compare them with other submillimeter detected, 
  infrared luminous galaxies in Section \ref{results}.
We discuss and summarize the results in Section \ref{discuss}.
Throughout,
  we adopt flat $\Lambda$CDM cosmological parameters:
  $H_0 = 70$ km s$^{-1}$ Mpc$^{-1}$, 
  $\Omega_{\Lambda}=0.7$ and $\Omega_{m}=0.3$.

\section{DATA}\label{data}

\subsection{Sample}\label{sample}

HG13 identified 47 local DOGs with
  $S_\nu$(12 $\mu$m) $>$ 20 mJy at $0.05<z<0.08$ 
  in the SDSS DR7.
These DOGs have extreme flux density ratios 
  between mid-infrared and UV bands
  with $S_\nu$(12 $\mu$m)/$S_\nu$(0.22 $\mu$m) $\geq892$.
The infrared luminosities of the DOGs are in the range
 $3\times10^{10} <L_{\rm IR}/{\rm L}_\odot <7\times10^{11}$
 with a median $L_{\rm IR}$ of $2.1\times10^{11}$ (${\rm L}_\odot$).
These infrared luminosities are based on an SED fit
  to the photometric data  at 6 $\mu$m $<\lambda\leq 140$ $\mu$m
  with the SED templates and fitting routine of \citet{mul11agn}, 
  DECOMPIR\footnote{http://sites.google.com/site/decompir}.
From these SED fits, 
  we computed the expected 880 $\mu$m flux densities for the 47 DOGs,
  and selected the four DOGs with the largest, predicted flux densities
  at 880 $\mu$m for SMA observation
  (see the target list in Table \ref{tab-samp}).

\subsection{Observations and Data Reduction}\label{obs}

Four local DOGs were observed in the compact configuration (8--70 m baselines)
  of the 8--element Submillimeter Array (SMA; \citealt{ho04}) interferometer at
  Mauna Kea, Hawaii in early 2013 (see Table \ref{tab-samp} for an observing journal).  
The SMA dual-sideband receivers were tuned to a local oscillator (LO) frequency of
  342 GHz (877 $\mu$m), and the correlator was configured 
  to process $2\times2$ GHz (intermediate frequency) IF bands 
  per sideband centered $\pm$4--8 GHz from the LO, 
  divided into 48 spectral ``chunks'' that each contained 64 individual
  1.6875 MHz channels.  
In each track, observations of two target DOGs were
  interleaved with nearby quasars on a 15 minute cycle.  
Additional observations of 3C 84, 3C 279, Uranus, and Titan were made 
  for calibration purposes when the science targets were at low elevations.  
Observing conditions were good, with precipitable water vapor levels at 1.5--2.0 mm 
  and stable phase behavior.

The raw visibilities were reduced with the {\tt MIR} software package.  
The bandpass response was calibrated with observations of 3C 84 and 3C 279, 
  and the antenna-based complex gains were determined by frequent observations of a
  nearby quasar: 0854+201 for LDOG-07, J1310+323 for LDOG-26, J1635+381 for
  LDOG-39, and L1549+026 for LDOG-41.  
The absolute amplitude scale was set based on observations of Uranus and Titan, 
  and should have a systematic uncertainty of $\sim$10\%\ or less.  
After calibration, the individual spectral channels  for each sideband and IF band 
  were combined into a composite wideband continuum visibility set.  
Those data were then Fourier inverted assuming natural weighting, 
  deconvolved with the {\tt CLEAN} algorithm, and then restored with a
  synthesized beam (with a FWHM of roughly $2\farcs2\times1\farcs8$).  
The imaging and deconvolution procedures were conducted with the {\tt MIRIAD}
  software package.  

\begin{figure*}
\center
\includegraphics[width=160mm]{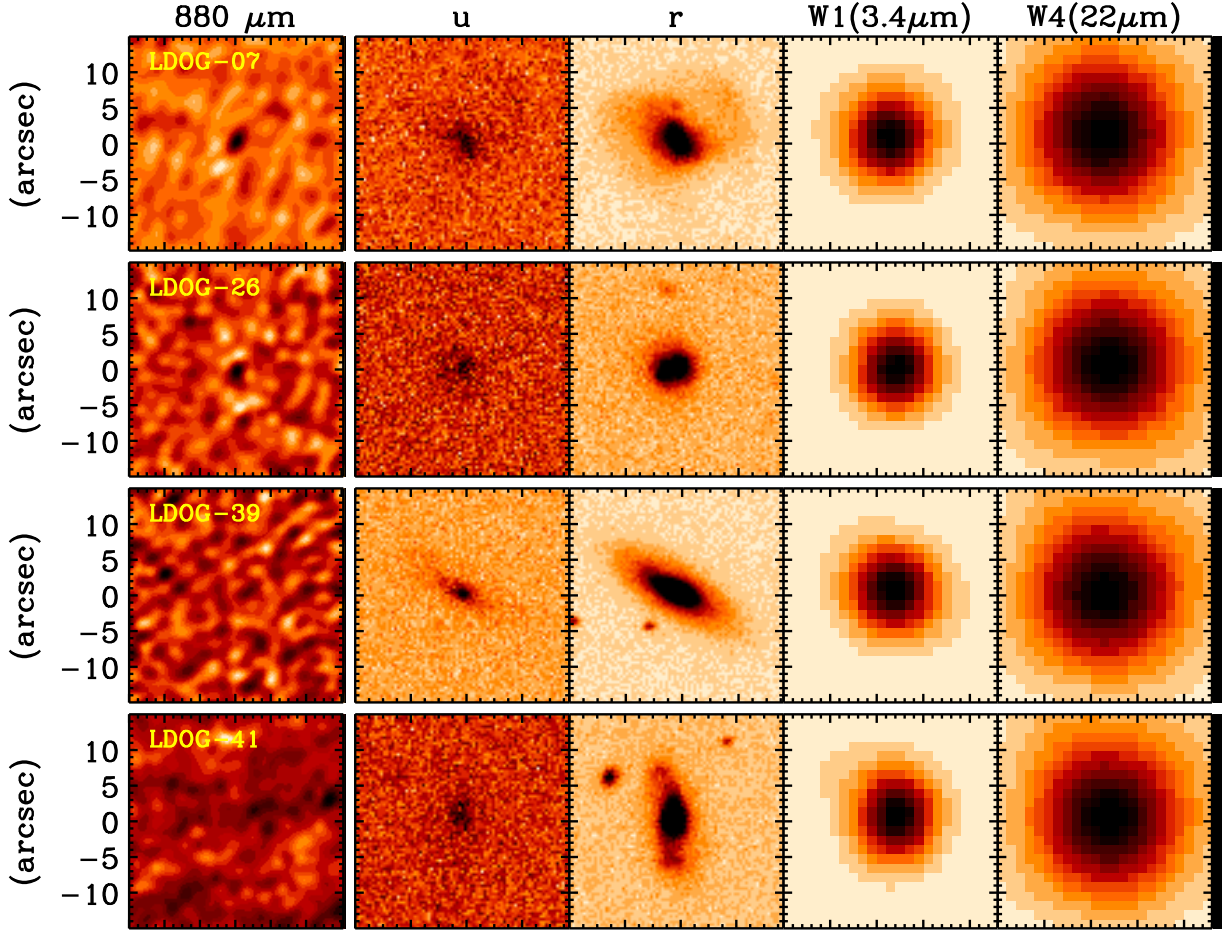}
\caption{SMA synthesis maps ($30\arcsec \times 30\arcsec$) of 
  the 880 $\mu$m continuum emission, and
  SDSS $ur$ and \wise 3.4/22 $\mu$m cutout images
  for the target DOGs.
The north is up, and the east is to the left.
}\label{fig-cut}
\end{figure*}

We show the resulting 880 $\mu$m continuum aperture synthesis images 
  for the four local DOG targets in Figure \ref{fig-cut};
  we also show SDSS $ur$ and \wise 3.4/22 $\mu$m cutout images.
None of the DOGs are resolved in the 880 $\mu$m synthesis images.
The SMA synthesis maps for two DOGs 
  in the top panels (LDOG$-$07 and LDOG$-$25) show clear detections
  with $S_\nu$(880 $\mu$m) $=10-13$ mJy and $S/N>5$.
However, the other two DOGs in the bottom panels (LDOG$-$39 and LDOG$-$41)
  are not visible in the synthesis maps.
They are are not detected with 3$\sigma$ upper limits of $S_\nu$(880 $\mu$m) $=5-9$ mJy.
We list the four target DOGs in Table \ref{tab-samp}
  with the SMA observation log
  and the measured 880 $\mu$m flux densities.

\section{RESULTS}\label{results}

\subsection{Determination of Physical Parameters of Dust Content in Local DOGs}\label{fit}

We first compute the infrared luminosities of the DOGs
  using the SED templates and fitting routine of \citet{mul11agn}.
This routine decomposes the observed SED of a galaxy into 
  two components (i.e., a host-galaxy and an AGN).
Therefore, we can also measure the contribution of (buried) AGN 
  to the total infrared luminosity of a galaxy.
This method is the same as in HG13,
  but we have additional submillimeter data 
  to constrain the fit.

The SED fit with the \citet{mul11agn} routine does not provide
  the dust temperatures and masses for the galaxies.
We thus fit the observational data again
  using a modified blackbody function
  with two (warm and cold) dust components 
  \citep{de01,vla05,wil09}:
    
\begin{equation}\label{eq-mbb}
 S_{\nu_{\rm obs}} = (1+z)   
   [A_w \nu_{\rm rest}^\beta B(\nu_{\rm rest},T_w) +  
    A_c \nu_{\rm rest}^\beta B(\nu_{\rm rest},T_c)],
\end{equation} 

where  $A_w$ and $A_c$ are the relative contributions
  of warm and cold dust components,
  $T_w$ and $T_c$ are dust temperatures,
  $B$($\nu$,$T$) is the Planck function, and
  $\beta$ is the dust emissivity index.
We examined two values of $\beta$ (i.e., 1.5 and 2.0),
  and found that $\beta=2.0$ generally provides better fits.
Therefore, we use $\beta=2.0$ for the fit,
  consistent with \citet{vla05} and \citet{wil09}.

We then compute the dust mass from the observed
  flux density \citep{hil83}, defined by 
\begin{eqnarray}
M_{\rm dust} & = & M_{\rm dust,w} + M_{\rm dust,c} \nonumber \\
 & =&  \frac{1}{1+z} \frac { D_L^2 } {k_d^{\rm rest}}  \left\lbrack  
  \frac {S_{\nu_{\rm obs}, w}} {B(\nu_{\rm rest},T_w)} + 
  \frac {S_{\nu_{\rm obs}, c}} {B(\nu_{\rm rest},T_c)}
  \right\rbrack  \nonumber \\
 & =&  \frac{1}{1+z} \frac { S_{\nu_{\rm obs}} D_L^2 } {k_d^{\rm rest}} \left\lbrack  
   \frac { A_w + A_c} {A_w B(\nu_{\rm rest},T_w) + A_c B(\nu_{\rm rest},T_c)}
   \right\rbrack ,
\label{eq-mass}
\end{eqnarray}

\begin{figure*}
\center
\includegraphics[width=155mm]{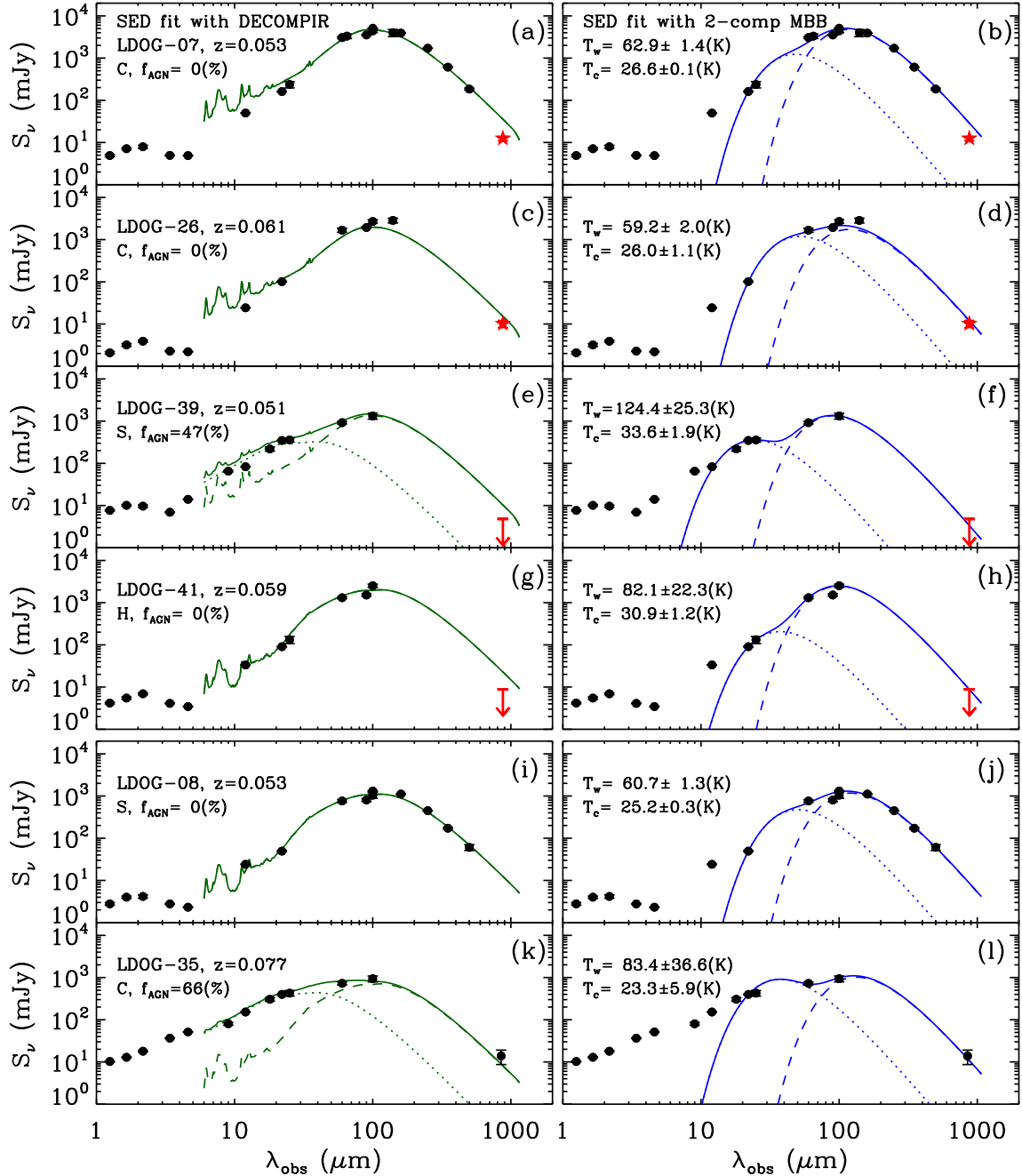}
\caption{SEDs of four DOGs with SMA observations (a--h),
  and of two DOGs with submillimeter data
  in the literature (i--l).
Red stars are SMA 880 $\mu$m data,
  and down arrows are upper limits.
Black filled circles are photometric data compiled in HG13.
There are error bars for all the points;
  they are mostly smaller than the symbols.
In the left panels,
  solid, dotted, and dashed lines indicate the best-fit SEDs 
  with the DECOMPIR routine of \citet{mul11agn}
  for total, AGN, and host-galaxy components, respectively.
Galaxy classification based on optical line ratios
  (H: SF, C: Composite, S: Seyfert)
  and the AGN contribution to the total infrared luminosity
  are shown in the top of each panel.
In the right panels,
  solid, dotted, and dashed lines indicate the best-fit SEDs 
  with the two-component modified blackbody function
  for total, warm and cold dust components, respectively.
Both fits use the data at 20 $\mu$m $<\lambda\leq 880$ $\mu$m. 
}\label{fig-sed}
\end{figure*}

where $k_d$ is the dust mass opacity coefficient, 
  $D_L$ is the luminosity distance, and 
  $S_{\nu_{\rm obs}}$ is the observed flux density\footnote{Note that
  if we use $S_{\nu_{\rm obs}}$ derived from equation (\ref{eq-mbb})
  instead of the observed flux density,
  equation (\ref{eq-mass}) can be simply expressed as
  $D_L^2 \nu_{\rm rest}^\beta (A_w + A_c)$ / $k_d^{\rm rest}$.}
  with $S_{\nu_{\rm obs}}=S_{\nu_{\rm obs},w} + S_{\nu_{\rm obs},c}$.  
We adopt $k_d^{\rm rest} = 0.383$ cm$^2$ g$^{-1}$
  at 850 $\mu$m from \citet{dra03}.
We use $k_d^{\rm rest}$ at 850 $\mu$m rather than at 880 $\mu$m
  to be consistent with the comparison sample of galaxies (see Section \ref{comp}).
Note that the $k_d$ value is usually very uncertain;
  it can change by a factor of 2 
  (e.g., $k_d^{850} = 0.77$ cm$^2$ g$^{-1}$ in \citealt{jam02}).
Therefore, the resulting dust mass can also change depending
  on the $k_d$ value adopted.
For $S_{\nu_{\rm obs}}$, 
  we use the flux densities expected from the modified blackbody fit
  at 850(1+$z$) $\mu$m.

We use the photometric data at $20~\mu$m $<\lambda\leq 880~\mu$m
  for both SED fits.
We also compile the far-infrared/submillimeter data in the literature,
  and include them for the fit
  (e.g., {\it Herschel} 100--500 $\mu$m data for LDOG$-$07
  from the H-ATLAS program; \citealt{rig11,pil10}).

\begin{deluxetable*}{cccccr}
\tabletypesize{\normalsize}
\tablewidth{0pc}
\tablecaption{SED Fit Parameters for Local DOGs
\label{tab-sed}}
\tablehead{
ID & L$_{\rm IR}$ & $T_{\rm warm}$    & $T_{\rm cold}$ & $M_{\rm dust}$            &  $M_{\rm cold}/M_{\rm warm}$  \\
   & ($\times10^{11}$L$_\odot$) & (K) & (K)            & ($\times10^{7}$M$_\odot$) &                
}
\startdata
LDOG-07 &  4.39$\pm$ 0.02 &  62.9$\pm$  1.4 &  26.6$\pm$  0.1 & 24.53$\pm$ 0.38 &             287 \\
LDOG-26 &  2.50$\pm$ 0.05 &  59.2$\pm$  2.0 &  26.0$\pm$  1.1 & 13.49$\pm$ 2.46 &              89 \\
LDOG-39 &         $<$2.14 & 124.4$\pm$ 25.3 &         $>$33.6 &         $<$1.91 &         $<$2592 \\
LDOG-41 &         $<$2.67 &  82.1$\pm$ 22.3 &         $>$30.9 &         $<$7.34 &         $<$1590 \\
\hline
LDOG-08 &  1.19$\pm$ 0.01 &  60.7$\pm$  1.3 &  25.2$\pm$  0.3 &  7.85$\pm$ 0.26 &             197 \\
LDOG-35 &  4.90$\pm$ 0.23 &  83.4$\pm$ 36.6 &  23.3$\pm$  5.9 & 21.41$\pm$14.72 &             660   
\enddata
\end{deluxetable*}

Figure \ref{fig-sed} shows the photometric data for the DOGs
  along with the best-fit SEDs for infrared luminosities (left panels)
  and for dust temperatures and masses (right panels).  
The SEDs for two DOGs in the middle panels (e--h)
  are not well constrained because the SMA flux densities
  are upper limits, not used for the SED fit.
We thus flag the derived quantities for these DOGs
  with lower and upper limits
  depending on parameters in the following Figures and Tables.

Table \ref{tab-sed} lists
  the infrared luminosities, dust temperatures
  of the warm and cold components,
  total dust masses, and
  dust mass ratios between warm and cold components
  of the four DOGs.
We compute the uncertainty in each parameter
  by randomly selecting flux densities
  at each band within the associated error distribution 
  (assumed to be Gaussian) and then refitting.

We compiled the far-infrared/submillimeter data in the literature, and
  found two more DOGs with existing submillimeter data.
LDOG$-$08 in Figure \ref{fig-sed}(i--j) has {\it Herschel} 100$-$500 $\mu$m data 
  from the H-ATLAS program \citep{rig11}, and
LDOG$-$35 in Figure \ref{fig-sed}(k--l) has SCUBA 850 $\mu$m data
  from \citet{ant04}.
We show these two DOGs in Figure \ref{fig-sed} and Table \ref{tab-sed},
  and include in our analysis.

\subsection{Comparison of Dust Content between
  Local DOGs and Infrared Luminous Galaxies with Submillimeter Detection}

To see whether the local DOGs with submillimeter detection
  are a population distinct from 
  other submillimeter detected, infrared luminous galaxies,
  we compare the dust parameters between the two populations.
Because of the small number of local DOGs with submillimeter data
  and because of the inhomogenous selection criteria for
  submillimeter detected, infrared luminous galaxies (see next Section),
  we simply examine their relative distribution
  in several parameter spaces.

\subsubsection{Local Infrared Luminous Galaxies with Submillimeter Detection}\label{comp}

Among many studies based on submillimeter observations of local galaxies
 (e.g., \citealt{wil09,cle10,dale12}),
 we select a comparison sample including
 only the galaxies with infrared luminosities similar to the local DOGs
 (i.e., $10^{11}   \lesssim L_{\rm IR}/{\rm L}_\odot \lesssim 10^{12}$) 
 and with submillimeter data at $\lambda\geq850$ $\mu$m.

We first use the galaxies 
  in the SCUBA local universe galaxy survey (SLUGS; \citealt{dun00, de01}).
Among 104 galaxies with SCUBA 850 $\mu$m data in the survey,
  we select 63 galaxies at $z>0.01$ with available mid- and far-infrared data.
The lower redshift limit removes very nearby, extended galaxies
  that could be resolved in the mid-infrared.  
For 48 out of 63 galaxies, 
  we use {\it WISE}, {\it IRAS}, and {\it AKARI} data at 3.4--160 $\mu$m
  from the SDSS galaxy catalog with multiwavelength data compiled in HG13.
For the remaining 15 galaxies,
  we adopt the mid- and far-infrared data
  from the Great Observatories All-sky LIRG Survey (GOALS; \citealt{arm09});
  {\it Spitzer} and {\it IRAS} data at 3.6--160 $\mu$m in \citet{u12}.
There could be some potential DOG candidates in this GOALS sample,
  not covered in HG13 (i.e., SDSS).
We do identify eight potential DOG candidates
  with $S_{\rm Spitzer~8\mu m}/S_{\rm GALEX~0.22\mu m}>982$
  in the GOALS sample, and 
  do not include them in the comparison sample\footnote{
  Note that the {\it GALEX} 0.22 $\mu$m and {\it Spitzer} 8 $\mu$m data
  are exactly equivalent to
  the $R$-band (0.65 $\mu$m) and \spitzer 24 $\mu$m data 
  originally used for selecting $z\sim2$ DOGs \citep{dey08}.}.
We do not include them in the DOG sample either,
  because the selection criteria (e.g., observed bands, 
  mid-infrared flux density limits) are not exactly the same
  as HG13.

We also use the luminous infrared galaxies with SMA 880 $\mu$m data
  in \citet{wil08}.
Among 15 galaxies in the paper,
  we include seven systems
  that do not overlap with the SLUGS sample and
  that do not have two distinct interacting galaxies.
Their mid- and far-infrared flux densities are again adopted
  from HG13 (six galaxies) and GOALS (one galaxy).

In summary, there are 62 galaxies
  with submillimeter, mid- and far-infrared data at $z>0.01$
  for comparison with the local DOGs.
We apply the same SED fitting routines of Section \ref{fit}
  to these galaxies to derive physical parameters
  including infrared luminosity, dust mass and temperature.
  
Using this sample,
  we first confirm that our measurements
  agree well with previous measurements:
  the dust masses in \citet{dun00}, \citet{de01} and \citet{wil09},
  and the dust temperatures in \citet{wil09}.
Moreover, we note that
  there are recent sophisticated models that provide
  several dust parameters simultaneously
  from the SED fit (e.g., \citealt{dl07,dac08}; see also \citealt{wal11} for a review).
Because of the small number of bands 
  in the far-infrared/submillimeter regimes for the DOGs,
  we restrict our analysis to simple models
  (e.g., two-component modified blackbody function)
  rather than sophisticated ones that require many observational data points.
Our simple approach works well.
For example, the dust masses derived in this study
  for the galaxies in \citet{wil09}
  show excellent agreement with those based on the \citet{dl07} models.

We apply the same fitting routine both
  to local DOGs and to other infrared luminous galaxies 
  with submillimeter detections.
Thus, the comparison between the two suffers
  no bias resulting from different SED fitting methods.

\begin{figure}
\center
\includegraphics[width=85mm]{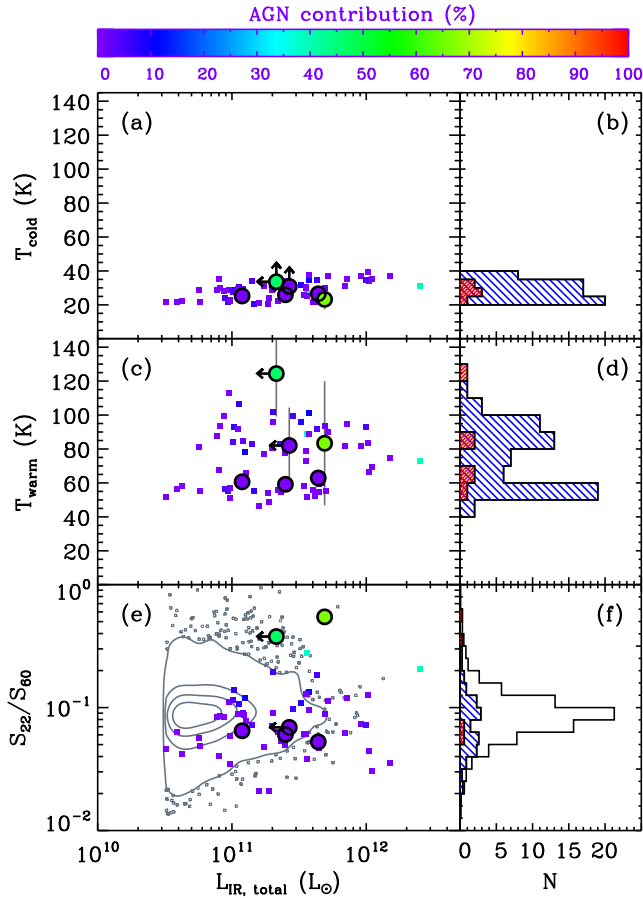}
\caption{Dust temperature $T_{\rm cold}$ of the cold component 
  for local DOGs (circles) and 
  for other infrared luminous galaxies with submillimeter detections (squares)
  as a function of total infrared luminosity (a), and their histograms (b).
Different colored symbols represent different 
  AGN contributions measured from the SED decomposition 
  (color coded as shown by the color bar to the top; see HG13 for details).
We plot error bars only for local DOGs.
DOGs and other infrared luminous galaxies are denoted by hatched histograms with
  orientation of 45$^\circ$ ($//$ with red color) and 
  of 315$^\circ$ ($\setminus\setminus$ with blue color) 
  relative to horizontal, respectively.
Same as (a--b), but for the dust temperature $T_{\rm warm}$ 
  of the warm component  (c--d) and
  for the flux density ratios between \iras 60 $\mu$m and \wise 22 $\mu$m (e--f).
The contours and gray dots in (e) indicate the distribution of
  {\it IRAS} 60 $\mu$m detected SDSS galaxies 
  at $z>0.01$ regardless of submillimeter detection.
The histograms in (f) are arbitrarily scaled down to match the range in other panels.
Left and up arrows indicate upper and lower limits, respectively.
}\label{fig-tdust}
\end{figure}

\subsubsection{Comparisons of Dust Temperature and Mass}

Figure \ref{fig-tdust} displays several parameters related to the dust temperature
  as a function of total infrared luminosity.
The top panels show the temperature $T_{\rm cold}$ of the cold dust component.
The cold dust temperature is in a very narrow range
  both for DOGs (circles) and for other infrared luminous galaxies (squares).
Remarkably, $T_{\rm cold}$ does not change much with infrared luminosity.

The temperature of the warm dust component, $T_{\rm warm}$, in the middle panel
  also does not depend on infrared luminosity,
  consistent with previous studies \citep{de01}.
However, it shows a large dispersion from 45 K to 125 K.
The dust temperatures of all the DOGs 
  except the one with $T_{\rm warm}\sim125$ K
  are well mixed with those of other infrared luminous galaxies 
  with similar infrared luminosities.
   
To examine the behavior of the warm dust component in galaxies,
  we plot the observed flux density ratio, $S_\nu$(22 $\mu$m)/$S_\nu$(60 $\mu$m),
  in the bottom panel.
For comparison, we also plot the contours and gray dots
 indicating the distribution of 
  {\it IRAS} 60 $\mu$m detected SDSS galaxies 
  at $z>0.01$ regardless of submillimeter detection.
The panel shows that the four DOGs with small AGN contribution (purple circles)
  are indistinguishable from other infrared luminous galaxies (squares).
Two DOGs and two infrared luminous galaxies with a large AGN contribution
  (green and cyan symbols)
  have larger flux density ratios than other galaxies,
  consistent with expectation \citep{deg85,vei09,leejc12akari}.

\begin{figure}
\center
\includegraphics[width=85mm]{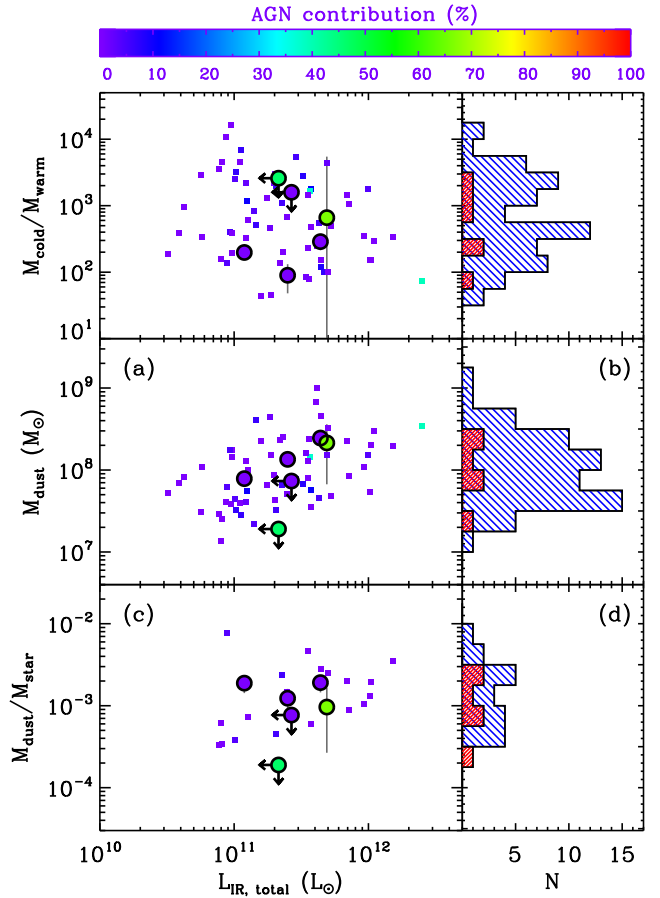}
\caption{Same as Figure \ref{fig-tdust}, but for
  the dust mass ratio between the cold and warm components (a--b),
  total dust mass (c--d), and
  mass ratio between the dust and stars in galaxies (e--f).
Left and down arrows indicate upper limits.
}\label{fig-mdust}
\end{figure}

In Figure \ref{fig-mdust}, we plot several parameters
  related to the dust mass as a function of infrared luminosity.
The top panels show the dust mass ratios
  between the cold and warm components.
Again the DOGs do not differ from other infrared luminous galaxies.  
We run a Kolmogorov-Smirnov (K-S) test to determine
  whether the DOGs (circles) and other infrared luminous galaxies (squares) 
  are drawn from the same distribution.
The K-S test cannot reject the hypothesis that
  the ratio distributions of the two samples are extracted from
  the same parent population.
If we run the K-S test for
  the galaxies in the same infrared luminosity range
  (i.e., $1.2\times10^{11}<L_{\rm IR, total}/L_\odot <4.7\times10^{11}$),
  the conclusion does not change.

The middle panels show the total dust mass, 
 $M_{\rm dust} = M_{\rm cold} + M_{\rm warm}$.
The dust mass roughly correlates with 
  infrared luminosity, consistent with previous studies \citep{de01,mag12}.
The Spearman correlation coefficient
  $\mathbf{(\rho_s)}$ is 0.51 and the probability of obtaining the correlation 
  by chance is $<$0.01\%,
  confirming the correlation between the two.
The dust masses of the DOGs
  are indistinguishable from other infrared luminous galaxies.
The K-S test also confirms this impression.

The bottom panels display the ratios between dust masses ($M_{\rm dust}$) 
  and stellar masses ($M_{\rm star}$).
We use the stellar mass estimates in 
  the MPA/JHU DR7 value-added galaxy 
  catalog\footnote{http://www.mpa-garching.mpg.de/SDSS/DR7/Data/stellarmass.html}.
These estimates are based on the fit to SDSS five-band photometry 
  with the \citet{bc03} models (see also \citealt{kau03}).
We convert the stellar masses in the catalog
  that are based on the Kroupa initial mass function (IMF; \citealt{kro01})
  to those with a Salpeter IMF \citep{sal55} 
  by dividing them by 0.7 \citep{elb07}.
The stellar masses in this catalog
  are not available for all the galaxies in this study.
If we use the stellar masses derived from {\it WISE} 3.4 $\mu$m luminosities
  to increase the sample size \citep{hwa12shels},
  the conclusions do not change.

The ratios, $M_{\rm dust}$/$M_{\rm star}$, for the majority of 
  the galaxy samples
  are between 10$^{-4}$ and 10$^{-2}$,
  consistent with previous results for star-forming galaxies in the local universe
  \citep{san10,dun11,ski11}.
The ratios for the DOGs except the AGN-dominated outlier again
  do not differ from other infrared luminous galaxies.

\section{DISCUSSION AND SUMMARY}\label{discuss}

We conducted SMA observations of four local analogs of DOGs
  to measure the physical parameters of their dust content.
Two DOGs are clearly detected at 880 $\mu$m with
 $S_\nu$(880 $\mu$m) $=10-13$ mJy and $S/N>5$;
  the other two are not detected 
  with 3$\sigma$ upper limits of $S_\nu$(880 $\mu$m) $=5-9$ mJy.
In addition to these four DOGs,
  we compiled submillimeter data for additional two DOGs from the literature.
Thus, we determine the dust temperatures and masses 
  for a total of six local DOGs.
The comparison of these DOGs with other infrared luminous galaxies 
  with submillimeter detection
  indicates no significant difference in dust parameters
  between the two populations.
  
Previous studies suggest that 
  there are two types of DOGs 
  for both local and high-$z$ DOGs:
  star formation (SF)- and AGN-dominated ones
  in their near- and mid-infrared SEDs (\citealt{dey08}; HG13).
The reason for the extreme flux density ratios
  between mid-infrared and UV bands in SF-dominated DOGs 
  mainly results from abnormal faintness in the UV rather
  than extreme brightness in the mid-infrared (\citealt{pen12}; HG13).
This conclusion also applies to AGN-dominated DOGs,
  but the large mid-infrared fluxes from the AGN dust also contribute to
  the extreme flux density ratios.

For the six local DOGs, 
  the dust masses and temperatures
  are similar to those of other submillimeter detected,
  infrared luminous galaxies with 
  similar infrared luminosities.
Thus, the DOGs are not a distinctive population
  among dusty galaxies.
In other words, the main reason they are selected as DOGs
  is not an extremely large dust content,
  but simply results from a large dust obscuration along the light of sight.
This conclusion explains the significant fraction of local DOGs
  with highly inclined disks (see Figure 6 in HG13; 
  see also \citealt{kar12} for disk-dominated, high-$z$ DOGs).
Merging processes also change the dust geometry
  to favor large dust obscuration \citep{pen12}.
  
The DOGs with large AGN contribution clearly contain a
  hot dust component with $T\gtrsim80$ K
  (see middle panels in Figure \ref{fig-tdust}).
Although the galaxies with a large AGN contribution tend to be
  selected as DOGs because of their large mid-infrared fluxes 
  (see bottom left panel in Figure 9 of HG13),
  not all infrared luminous galaxies 
  with large AGN contribution are selected as DOGs.
    
One interesting feature of the AGN-dominated DOGs is that
  their cold temperatures are
  similar to those of SF-dominated DOGs and other infrared luminous galaxies
  (see top panels in Figure \ref{fig-tdust}).
This result is consistent with recent conclusions
  that the effect of AGN on star-forming galaxies does not
  appear on the `Rayleigh-Jeans' side of the infrared SED peak (i.e., cold components),
  but only appears on the `Wien' side (i.e., warm components) \citep{hat10,kir12}.

There are several studies on the dust temperatures and
  masses for high-$z$ DOGs \citep{buss09,mel12,saj12,wu12}.
Because of the different SED fitting methods and because of
  the small number of far-infrared/submillimeter data
  for high-$z$ DOGs,
  a direct comparison of dust parameters
  between local and high-$z$ DOGs is not very meaningful.
Moreover, the infrared luminosity range for high-$z$ DOGs
  does not overlap with local DOGs.

A rough comparison of the dust temperatures
  based on currently available data (see Figure 13 in HG13)
  suggests that the dust temperatures for the majority of
  high-$z$ DOGs are similar to or lower than for local DOGs
  even though the infrared luminosities of high-$z$ DOGs
  are much higher than for local DOGs.
There are also some {\it hot} DOGs at high redshift with dust temperatures
  much higher than for local DOGs \citep{wu12}.
Far-infrared and submillimeter data
  for a larger number of DOGs in both low and high redshifts
  with similar infrared luminosities
  will be useful for a thorough comparison between the two populations.
  
This study clearly shows the importance of submillimeter data
  in understanding the dust content of local DOGs.
We plan to extend this study to a larger sample of local DOGs
  with the Caltech Submillimeter Observatory (G.-H. Lee et al., in preparation).

{\it Facility:} \facility{SMA}

\acknowledgments

We thank the anonymous referee for his/her useful comments that improved the manuscript.
We also thank Jong Chul Lee, Gwang-Ho Lee and Jubee Sohn for useful discussions.
H.S.H. acknowledges the Smithsonian Institution for the support of his post-doctoral fellowship.
The Smithsonian Institution also supports the research of MJG.
The Submillimeter Array is a joint project between the Smithsonian Astrophysical Observatory 
and the Academia Sinica Institute of Astronomy and Astrophysics and 
is funded by the Smithsonian Institution and the Academia Sinica.
Herschel is an ESA space observatory with science instruments provided by 
European-led Principal Investigator consortia and with important participation from NASA.

\bibliographystyle{apj} 
\bibliography{ref_hshwang} 



\end{document}